# Pressure-induced lattice collapse in tetragonal phase and its equation of state at 300K in single crystalline $Fe_{1.05}Te$

Chao Zhang[1], Wei Yi[1], Liling Sun[1]*, Xiao-Jia Chen[2,3], Russell J. Hemley[2], Ho-kwang Mao[2], Wei Lu[1], Xiaoli Dong[1], Ligang Bai[4], Jing Liu[4], Antonio F. Moreira Dos Santos[5], Jamie J. Molaison[5], Christopher A. Tulk[5], Genfu Chen[1], Nanlin Wang,[1] and Zhongxian Zhao[1]*

[1]Institute of Physics and Beijing National Laboratory for Condensed Matter Physics, Chinese Academy of Sciences, Beijing 100190, P. R. China

[2]Geophysical Laboratory, Carnegie Institution of Washington, Washington, DC 20015, U.S.A

[3]Department of Physics, South China University of Technology, Guangzhou 510640, P. R. China

[4]Institute of High Energy Physics, Chinese Academy of Sciences, Beijing 100039, P. R. China

[5]Neutron Scattering Science Division, Oak Ridge National Laboratory, Oak Ridge, TN 37831, U.S.A.

Extensive measurements of X-ray diffraction, neutron diffraction, resistance, and magnetization were performed on single crystalline $Fe_{1.05}Te$ under pressure. A pressure-induced lattice collapse was observed in tetragonal phase at pressure of 4 GPa and at room temperature. The onset temperature of the structural phase transition was found to decrease with increasing pressure but increase upon further compression after passing through a minimum around the collapse. However, the onset of antiferromagnetic transition scarcely changes with pressure. No superconductivity was detected at pressures up to 20 GPa.

Corresponding author:
llsun@aphy.iphy.ac.cn
zhxzhao@aphy.iphy.ac.cn

The discovery of superconductivity with critical transition of 26 K in $LaFeAsO_{1-x}F_x$ [1] expedited the research for this new family of superconductor. Four types of materials, ReFeAsO (Re= Ce, Pr, Nd, Sm etc), $AFe_2As_2$ (A=Ba, Sr, Ca ), AFeAs (A=Li, Na) and FeSe(Te), have been discovered [2-4]. Superconductivity in the binary α-FeSe with transition temperature of 8 K was found [4] This compound has PbO-type tetrahedral structure, with stacking FeSe layers along the *c*-axis. Partial substitution of Te for Se [5] enhances the Tc to 10 K. This superconducting transition temperature was further increased by application of pressure, reaching 27 K [6] at 1.48 GPa and 36~37 K with increasing pressure [7-8]. The compound of α-FeTe has very similar structure to tetragonal FeSe at ambient pressure. Theoretical calculations [9] demonstrated that α-FeTe adopts multiple Fermi surfaces, similar to that of $ReFeAsO_{1-x}F_x$, with hole pockets at the zone center and electron pockets at the zone corner. The antiferromagnetic (AF) transition also can be observed in α-$Fe_{1+y}Te$, as ReFeAsO and $AFe_2As_2$ parent compound exhibited. Neutron scattering studies at ambient pressure [10-11] found that the in-plane spin structure of $Fe_{1+y}Te$ is completely different from that of ReFeAsO and $AFe_2As_2$ whose moments form a collinear antiferromagnetic structure with spin direction along the *a*-axis in FeAs layer. Upon decrease temperature down to 65 K, the lattice distortion drives the $Fe_{1+y}Te$ compound from the tetragonal (T) to the monoclinic (M) phase [10-11]. Since the non-superconducting parent compounds of LaFeAsO and $BaFe_2As_2$ showed superconductivity under high pressure [12-14], the reported results motivate this investigation to explore pressure effect on structural and magnetic transition as well as pressure-induced potential superconductivity in $Fe_{1.05}Te$. Here we report high-pressure synchrotron x-ray diffraction, neutron diffraction, resistance and magnetization studies on $Fe_{1.05}Te$ compound. A remarkable reduction in c axis has been observed. This is the first observation of pressure-induced lattice collapse in the tetragonal phase of $Fe_{1.05}Te$ at room temperature. The equations of state of T phase and cT phase were determined from neutron diffraction results. A phase diagram of pressure dependence of the onset temperature of structural transition was also established through resistance measurements.

High pressure was created using a diamond anvil cell. Diamonds with low birefringence were selected for the high-pressure XRD measurements. Diamond anvils used were cut with a 300 μm culet, with a 100 μm diameter sample hole in a stainless steel gasket. The single crystalline sample was loaded into the hole with silicon oil to maintain the sample in a hydrostatic pressure environment. Pressure was applied in a direction normal to the *a-b* plane of the sample and determined by ruby fluorescence method at room temperature [15]. Angle-dispersive XRD experiments were carried out at the Beijing Synchrotron Radiation Facility (BSRF). A monochromatic x-ray beam with a wavelength of 0.6199 Å was used for all XRD measurements. The XRD images were collected using a charge-coupled device (CCD) detector, and the XRD geometry was calibrated with $CeO_2$.

Two individual resistance measurements were carried out in a diamond anvil cell made of Be-Cu alloy. The four-standard-probe technique was adopted in the experiments, as reported in Ref. [16]. No pressure medium was used in high-pressure resistance measurements. The high-pressure resistance as a function of temperature was measured using a CSW-71 cryostat. The magnetization measurements under ambient pressure were carried out using Quantum Design Magnetic Property Measurement System (MPMS-XL1), and those under hydrostatic pressure were conducted using a commercial pressure cell ( Mcell 10) specialized for MPMS-XL1. The experimental details can be found in Ref. [17].

Time-of-flight neutron diffraction experiments were performed at the SNAP (Spallation Neutrons and Pressure) beamline of the Spallation Neutron Source at Oak Ridge National Laboratory. To obtain a set of diffraction peaks, the powder sample ground from the same single crystal was used for the measurements. Pressure was generated with diamond anvils in Paris-Edinburgh high-pressure cell. The sample was loaded into TiZr gasket hole with pressure medium of a 4:1mixture of methanol and ethanol. The neutron diffraction data were collected through three banks and averaged from them.

The sample was synthesized by solid state reaction at ambient pressure, as described in a previous report [18]. Characterization by x-ray diffraction indicated that the sample used in this study is single crystal with tetragonal symmetry, as shown in Fig.1 (a). Both of resistance and magnetization measurements at ambient pressure exhibited anomalies at the same temperature near 65 K, as shown in Fig.1 (b) to (c), consistent with our previous experimental results that the abrupt decrease in resistance and magnetization below 65 K is related to the first-order T-M phase transition and AF transition respectively.

Fig.2 shows x-ray diffraction (XRD) patterns of $Fe_{1.05}Te$ at ambient and high pressures at room temperature. Inasmuch as the wavelength of synchrotron x-ray beam employed for high-pressure experiments is different from the wavelength (1.54 Å) of Cu $K_{\alpha}$ radiation used for ambient pressure measurement, we applied d-spacing value in Fig.2 for comparison, instead of two theta. Only the (003) reflection of the single crystalline sample can be detected over the angular range available at the experimental diffraction conditions. No new peaks were found in the diffraction under pressure up to 11.5 GPa, however, remarkable shift of peak (003) was observed during compression to 4 GPa, as displayed in Fig.2 (a). High-pressure neutron diffraction (ND) experiments were carried out for the polycrystalline $Fe_{1.05}Te$ sample in order to further confirm the observed lattice distortion. The ND patterns were shown in Fig.2 (b). All peaks collected at 0.7 GPa can be well indexed as tetragonal form. It was found the tetragonal phase persists up to 7 GPa, entirely consistent with our XRD results. We computed lattice parameter $c$ and $a$ value on basis of ND patterns obtained at different pressures and summarized the data together with the $c$ value calculated from XRD pattern in Fig.3. Apparently, two sets of $c$ values derived from XRD and ND measurements are in good agreement. A large reduction in $c$ axis was observed under applied pressure, as shown in Fig.3(a). The $c$ value is reduced about 5% at ~4 GPa. The striking reduction in $c$ direction suggested that applied pressure drives a lattice collapse in the tetragonal phase of $Fe_{1.05}Te$. To distinguish pressure-induced collapsed tetragonal phase from the tetragonal (T) phase, here we defined the high-pressure phase as cT phase. Pressure dependence of parameter $a$ is plotted in Fig.3 (b). We noted that the response of $a$ to pressure is different from parameter $c$, $a$ value decreases with pressure linearly. The unit cell volumes (V) as a function of pressure (P) in the T and cT phase were plotted in Fig.4. No visible discontinuity was found, further supporting that no first-order phase transition occurred under high pressure at least to 7 GPa. The data was fitted by the third order Birch-Murnaghan equation of state [19]:

$$P = \frac{3}{2}B_0\left[(V/V_0)^{-7/3} - (V/V_0)^{-5/3}\right]\left[1 + \frac{3}{4}(B_0' - 4)((V/V_0)^{-2/3} - 1)\right]$$

Where $B_0$ is the isothermal bulk modulus at zero pressure, $B_0'$ is the pressure derivative of $B_0$ evaluated at zero pressure, and $V/V_0$ is the ratio of high-pressure volume and zero-pressure volume of the sample. The resulting parameters are listed in Table 1. We obtained $B_0$= 31.3±1.45 GPa, $B_0'$= 6.6±1.2 for the T phase and $B_0$= 86.7±6.6 GPa, $V_0$= 88.7±0.4 (Å)$^3$ for the cT phase when $B_0'$= 4 was fixed. These results provide the crucial information for future theoretical and experimental investigations of the high pressure-behavior of FeTe parent compound.

To probe the electronic properties in the T and cT phase, and track the evolution of the onset temperature of structural transition with pressure, two separate high-pressure resistance measurements were carried out in a diamond anvil cell. Fig. 5 shows the representative electrical resistance (R) of the $Fe_{1.05}Te$ sample as a function of temperature (T) under high pressure up to 20 GPa. It was seen that the breadth of the transition of the compressed sample became broadened with increasing pressure. To make the data comparable, we define the maximum of dR/dT as the onset transition temperature. The onset temperature of the phase transition is pressure sensitive, shifts to low temperature side during compression, as shown in Fig.5 (a). Interestingly, the onset temperature shifted back to higher temperature side at 4 GPa where is the critical pressure of the T-cT phase transition. This upward shift of the onset is obvious in the R-T plot as displayed in Fig.5 (b). With further applying pressure up to 14.7 GPa, the sample lost its metallic character, behaving like a semiconductor. This dramatic change in electronic properties may be related to an additional phase transition. Polycrystalline x-ray diffraction at higher pressure (~14 GPa) is necessary to clarify this phenomenon. No superconductivity was observed upon uploading, in agreement with high-pressure resistance measurements for $FeTe_{0.92}$ recently reported by Okada et al [20]. Downloading from the maximum pressure investigated down to 4.7 GPa, we found the sample still hold semiconductor feature, as illustrated in Fig.5 (c). One potential reason for the observed irreversibility is due to a hysteretic effect from the proposed additional phase transition at the pressure where the sample lost its metallic behavior, as mentioned above. Another reason may be from the stress effect on the sample which can not recover back during releasing pressure down to 4.7 GPa.

Inasmuch as the structural and magnetic transition in $Fe_{1.05}Te$ take place at the same temperature (65K), we can not conclusively determine if the change in the onset transition temperature with pressure corresponds to the onset of T-cT phase transition. In order to distinguish pressure effect on the onset of T-cT phase and AF transition separately, we performed magnetization measurements under pressure for the $Fe_{1.05}Te$ sample. Fig.6 shows the temperature dependence of magnetization under fixed pressures. It was seen that the onset temperature of the AF transition ($T_{AF}$) remains nearly unchanged within resolution with increment of pressure to 0.8 GPa which is the highest pressure (~1GPa) of our magnetization measurement available. Next, we compared pressure-induced R-T and M-T shift at the similar pressure level and found there is 6.3 K shift of R-T curve at 0.6 GPa while little shift of the M-T curve at 0.8 GPa, strongly implying that the onset detected by resistance measurements is assigned to structural phase transition rather than magnetic transition. Here we use $T_{STR}$ to represent the onset temperature of structural phase transition.

The $T_{SRT}$, together with resistance data obtained at 150 K, as a function of pressure were plotted in Fig.7. Clearly, a notable change can be seen upon traversing of the T-cT phase. Firstly, the pressure dependence of $T_{STR}$ suddenly changed at the T-cT phase boundary. The $T_{STR}$ decreases with the pressure in the T phase, and in turn increases in the cT phase. Secondly, the pressure dependence of the resistance has a similar sign change at the boundary. The collapse tetragonal phase was also found in $CaFe_2As_2$ under applied pressure to 2.5 GPa where *c* parameter is reduced about ~1 (Å) [21-22]. Theoretical calculation [23] and inelastic neutron scattering experimental results under pressure [24] explained this strange behavior by that strong suppression of Fe-spin state reduced Fe-As bonding and enhanced corresponding As-As bonding. A striking increase in conductivity of $CaFe_2As_2$ upon traversing from T phase to cT phase has been observed experimentally [25-28] which is consistent with our results in which the resistance decreases from T phase to cT phase with pressure. The enhancement of conductivity upon travelling from T phase to cT phase of $CaFe_2As_2$ is considered to be relevant to Fe-spin state in FeAs layer [25-28]. Weak spin state of Fe atom diminishes electron scattering which is directly responsible for the conductivity.

We found there is a different manner in the cT phase. The resistance increases with pressure. We consider this change may be attributed to impurity of excess Fe in $Fe_{1.05}Te$ compound. In fact, there is a competition of spin state between excess Fe and Fe in FeTe after the lattice collapse. Under high pressure, remarkable reduction in *c* axis also shortened the distance of the excess Fe atoms. It is most likely that the spin state of excess Fe was not suppressed by increment of pressure. In contrast, reduction in distance between excess Fe atoms enhances the electron scattering, which results in increase in resistivity. Further study is needed to identify these important problems, particularly neutron scattering experiments under high pressure and low temperature.

**Acknowledgements**

We acknowledge interesting discussions with Prof. Pengcheng Dai and Prof. Xi Dai. This work was supported by the NSFC Grants No. 10874230, 10874211, 10804127, and 10874046, by the 973 project and Chinese Academy of Sciences. The work at Carnegie was supported by the U.S. DOE-NNSA (DEFC03-03NA00144). SNAP was supported by the scientific user facilities division of the U.S. DOE-BES at the Spallation Neutron Source. We acknowledge the support from EU under the project CoMePhS.

## References

[1] Y. Kamihara, T. Watanabe, M. Hirano and H. Hosono, J. Am,Chem. Soc. **130**, 3296 (2008); Z. A. Ren, W. Lu, J. Yang, W. Yi, X. L. Shen, Z. C. Li, G. C. Che, X. L. Dong, L. L.Sun, F. Zhou and Z. X. Zhao, Chin. Phys. Lett. **25**, 2215 (2008); G. F. Chen, Z. Li, D. Wu, G. Li, W. Z. Hu, J. Dong, P. Zheng, J. L. Luo, N. L. Wang , Phys. Rev. Lett. **100**, 247002 (2008); X. H. Chen, T. Wu, G. Wu, R. H. Liu, H. Chen, D. F. Fang, Nature, **453**, 7196 (2008).

[2] M. Rotter, M. Tegel, I. Schellenberg, W. Hermes, R. Pöttgen, D. Johrendt, Phys. Rev. B **78**, 020503(R) (2008); K. Sasmal, B. Lv, B. Lorenz, A. Guloy, F. Chen, Y. Xue, C. W. Chu, Phys. Rev. Lett. **101**, 107007 (2008).

[3] X. C. Wang, Q. Q. Liu, Y. X. Lu, W. B. Gao, L. X. Yang, R. C. Yu, F. Y. Li, C. Q. Jin, arXiv: 0806.4688; D. R. Parker, M. J. Pitcher, P. J Baker, I. Franke, T. Lancaster, S. J. Blundell, S. J. Clarke,Chemical Communications, 2189 (2009) or arXiv: 0810.3214; M. J. Pitcher, D. R. Parker, P. Adamson, S. J. C. Herkelrath, A. T. Boothroyd, S. J. Clarke, Chemical Communications, 5918(2008) or arXiv: 0807.2228; C. W. Chu, F. Chen, M. Gooch, A. M. Guloy, B. Lorenz, B. Lv, K. Sasmal, Z. J. Tang, J. H. Tapp and Y. Y. Xue, arXiv: 0902.0806.

[4] F. C. Hsu, J. Y. Luo, K. W. Yeh, T. K. Chen, T. W. Huang, P. M. Wu, Y. C. Lee, Y. L. Huang, Y. Y. Chu, D. C. Yan and M. K. Wu, PNAS, **105**, 14262 (2008).

[5] M. H. Fang, L. Spinu, B. Qian, H. M. Pham, T. J. Liu, E. K. Vehstedt, Y. Liu and Z. Q. Mao, arXiv: 0807. 4775.

[6] Y. Mizuguchi, F. Tomioka, S. Tsuda, T. Yamaguchi, Y. Takano, Appl. Phys. Lett. **93**, 152505 (2008).

[7] S. Medvedev, T. M. McQueen, I. Trojan, T. Palasyuk, M. I. Eremets, R.J. Cava,

S. Naghavi, F. Casper, V. Ksenofontov, G. Wortmann, C. Felser, arXiv: 0903.2143.

[8] S. Margadonna, Y. Takabayashi, Y. Ohishi, Y. Mizuguchi, Y. Takano, T. Kagayama, T. Nakagawa, M. Takata, and K. Prassides, arXiv:0903.2204.

[9] A. Subedi, L. Zhang, D. J. Singh, M. H. Du. Phys. Rev. B **78**, 134514 (2008).

[10] S. L. Li, C. Cruz, Q. Huang, Y. Chen, J. W. Lynn, J. P. Hu, Y. L. Huang, F. C. Hsu, K.W. Yeh, M. K. Wu, P. C Dai, Phys. Rev. B **79**, 054503 (2009).

[11] W. Bao, Y. Qiu, Q. Huang, M.A. Green, P. Zajdel, M. R. Fitzsimmons, M. Zhernenkov, M. Fang, B. Qian, E. K. Vehstedt, J. Yang, H. M. Pham, L. Spinu, Z.Q. Mao, arXiv: 0809.2058.

[12] H. Okada, K. Igawa, H. Takahashi, Y. Kamihara, M. Hirano, H. Hosono, K. Matsubayashi and Y. Uwatoko, arXiv: 0810.1153 or to be published in J. Phys. Soc. Jpn. **77** No.11 (2009).

[13] A. Mani,N. Ghosh,S.Paulraj,A. Bharathi, C. S. Sundar arXiv: 0903.4236.

[14] P. L. Alireza, Y. T. Chris Ko, J. Gillett, C. M. Petrone, J. M. Cole, S.E. Sebastian1 and G. G. Lonzarich. J. Phys.: Condens. Matt. **21**, 012208 (2009).

[15] H. K. Mao and P. M. Bell, Rev. Sci. Instrum., **52**, 615 (1981).

[16] W. Yi, L. L. Sun, Z. A. Ren, X. L. Dong, H. J. Zhang, X. Dai, Z. Fang, Z. C. Li, G. C. Che, J. Yang, X. L. Shen, F. Zhou and Z. X. Zhao, Europhys. Lett., **83**, 57002 (2008).

[17] W. Lu, J. Yang, X. L. Dong, Z. A. Ren, G. C. Che. and Z. X, Zhao , New J. Phys. **10**, 063026 (2008).

[18] G. F. Chen, Z, G. Chen, J. Dong, W. Z. Hu, G. Li, X. D. Zhang, P. Zheng, J. L. Juo and N. L. Wang, Phys. Rev. B **79**, 140509(R) (2009).

[19] F. Birch, J. Geophys. Res. **83**, 1257 (1978).

[20] H. Okada, H. Takahashi, Y. Mizuguchi, Y. Takano, and H Takahashi, arXiv: 0904.2945.

[21] A. I. Goldman, A. Kreyssig, K. Prokeš, D. K. Pratt, D. N. Argyriou, J. W. Lynn, S. Nandi, S. A. J. Kimber, Y. Chen, Y. B. Lee, G. Samolyuk, J. B. Leão, S. J. Poulton, S. L. Bud'ko, N. Ni, P. C. Canfield, B. N. Harmon, and R. J. McQueeney, Phys. Rev. B **79**, 024513 (2009).

[22] A. Kreyssig, M. A. Green, Y. Lee, G. D. Samolyuk, P. Zajdel, J. W. Lynn, S. L. Bud'ko, M. S. Torikachvili, N. Ni, S. Nandi, J. B. Leão, S. J. Poulton, D. N. Argyriou, B. N. Harmon, R. J. McQueeney, P. C. Canfield, and A. I. Goldman, Phys. Rev. B **78**, 184517 (2008).

[23] T. Yildirim, Phys. Rev. Lett. **102**, 037003 (2009).

[24] D. K. Pratt, Y. Zhao, S. A. J. Kimber, A. Hiess, D. N. Argyriou, C. Broholm, A. Kreyssig, S. Nandi, S. L. Bud'ko, N. Ni, P. C. Canfield, R. J. McQueeney, and A. I. Goldman, Phys. Rev. B **79**, 060510(R) (2009).

[25] H. Lee, E. Park, T. Park, F. Ronning, E. D. Bauer and J. D. Thompson, arXiv: 0809. 3550.

[26] M. S. Torikachvili, S. L. Bud'ko, N. Ni, and P. C. Canfield, Phys. Rev. Lett. **101**, 057006 (2008).

[27] T. Park, E. Park, H. Lee, T. Klimczuk, E. D. Bauer, F. Ronning, and J. D. Thompson, J. Phys.: Condens. Matter **20**, 322204 (2008).

[28] W. Yu, A. A. Aczel, T. J. Williams, S. L. Bud'ko, N. Ni, P. C. Canfield, and G. M. Luke, Phys. Rev. B **79**, 020511(R) (2009).

## Figure captions

**Fig. 1** (a) X-ray diffraction pattern of single crystalline $Fe_{1.05}Te$ obtained at ambient pressure and room temperature with Cu $K_{\alpha}$ radiation. (b) and (c) resistance and magnetization as a function of temperature of the sample at ambient pressure, showing structural (T-M) phase transition and antiferromagnetic transition at same temperature near 65 K.

**Fig. 2** (a) Representative high-pressure XRD patterns of $Fe_{1.05}Te$ obtained with a monochromatic beam ($\lambda$ = 0.6199 Å) at 300K; stars * indicate diffraction peaks from the metal gasket; labeled peak is from the sample. (b) high-pressure ND patterns of the polycrystalline sample at different pressures.

**Fig. 3** Pressure dependence of lattice parameter $c/c_0$ (a) and $a/a_0$ (b), showing a large lattice distortion in *c* axis in T phase.

**Fig. 4** Unit cell volume of T phase and cT phase as a function of pressure; solid circles, experimental data; curves, third order Birch-Murnaghan equation of state fit to the data obtained from ND measurements.

**Fig. 5** Resistance (R) versus temperature (T) at different pressures. (a) R-T curves shift to lower temperature side with pressure. (b) Inverse shift of R-T curves with increasing pressure. (d) R-T curves upon downloading from the maximum pressure down to 4.7 GPa. The inset of the figure (b) shows the definition of onset transition temperature.

**Fig. 6** Temperature dependence of magnetization of $Fe_{1.05}Te$ at four pressures (ambient, 0.1, 0.5 and 0.8 GPa), showing onset $T_{AF}$ remains nearly unchanged with applied pressure up to 0.8 GPa under 1 Tesla.

**Fig. 7** Pressure-temperature phase diagram constructed from resistance, XRD and ND

measurements. The red solid circle represents $T_{STR}$ and the purple represents the $R_{150K}$.

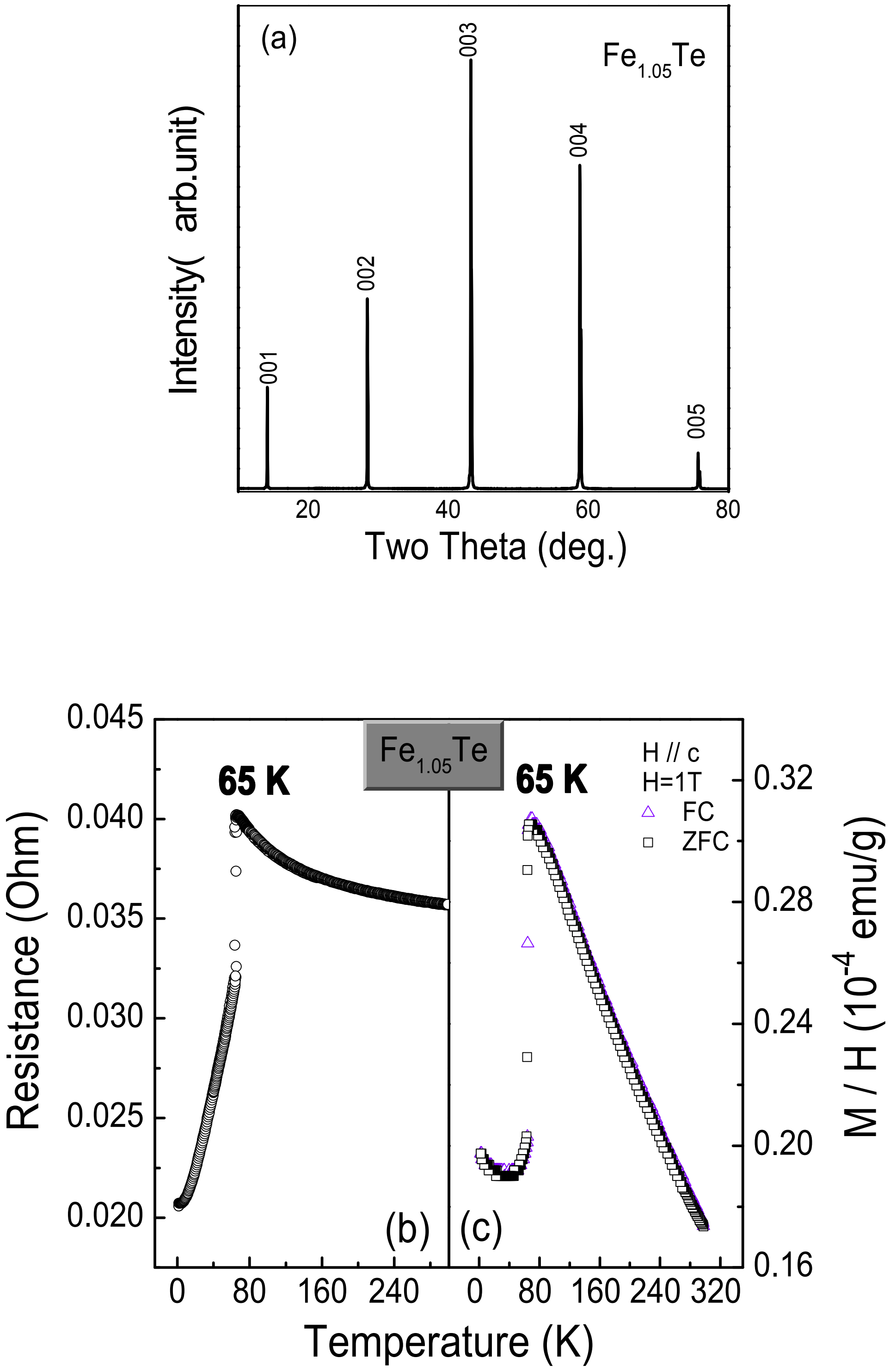

Fig. 1

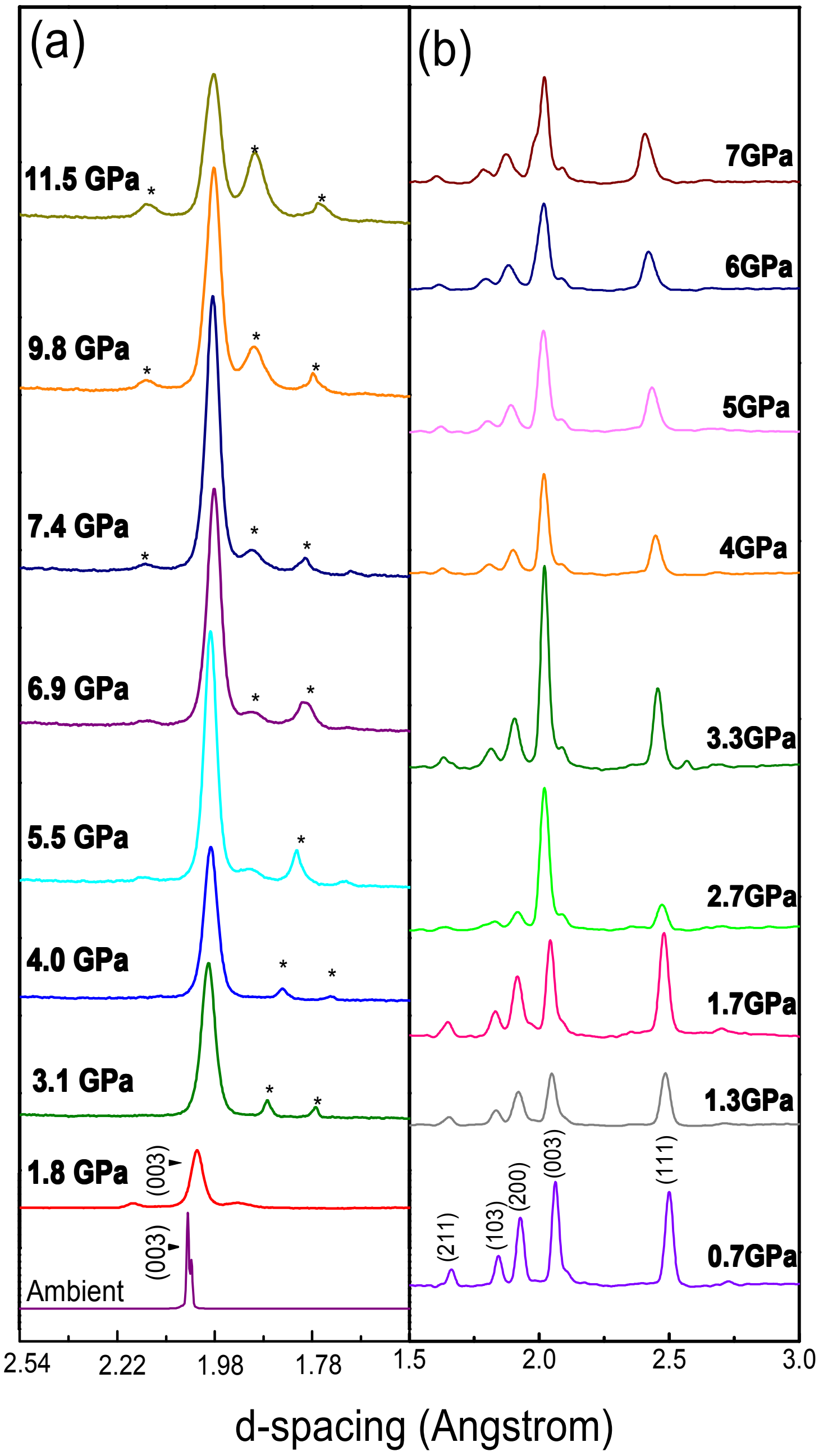

Fig.2

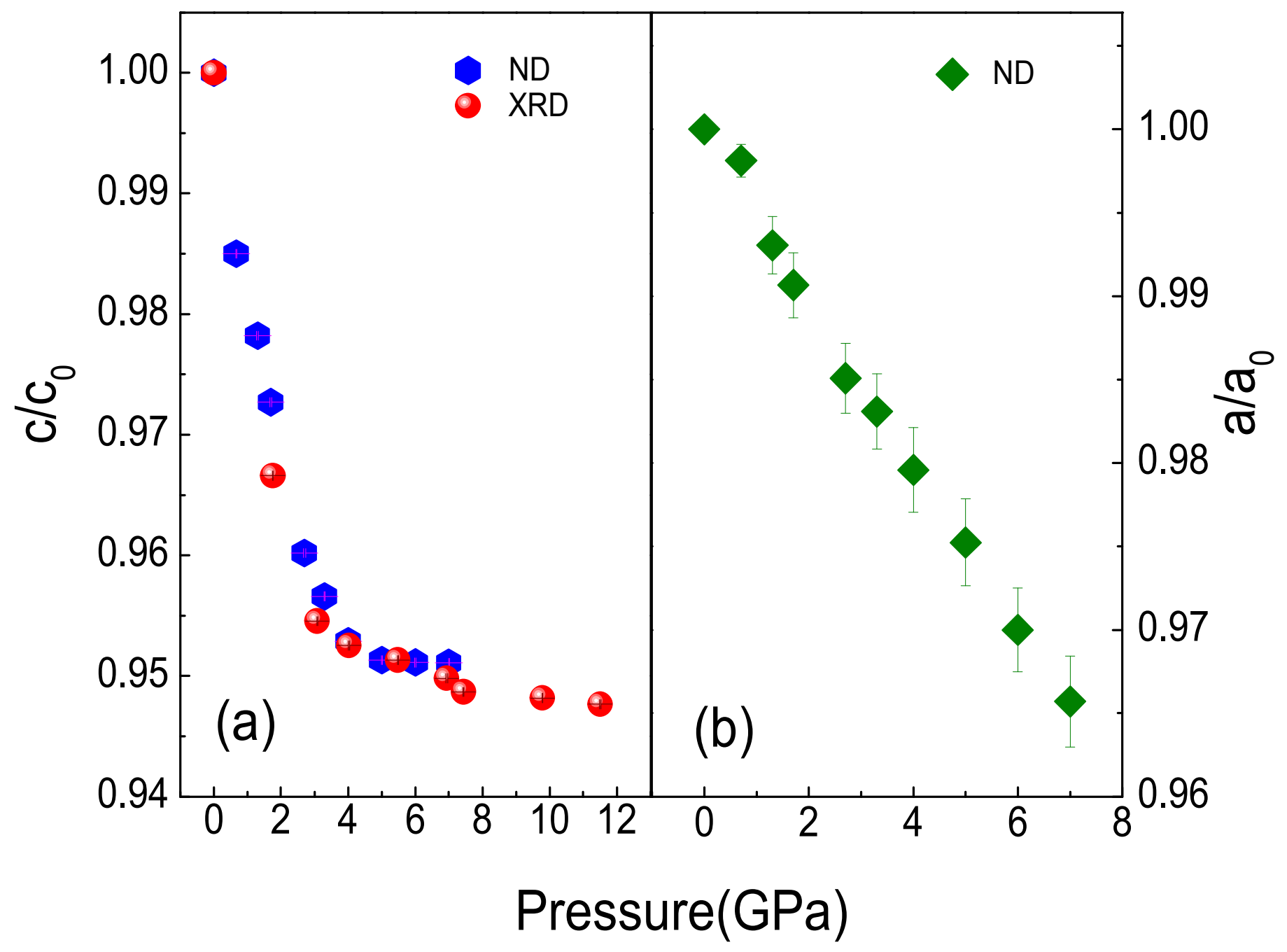

Fig. 3

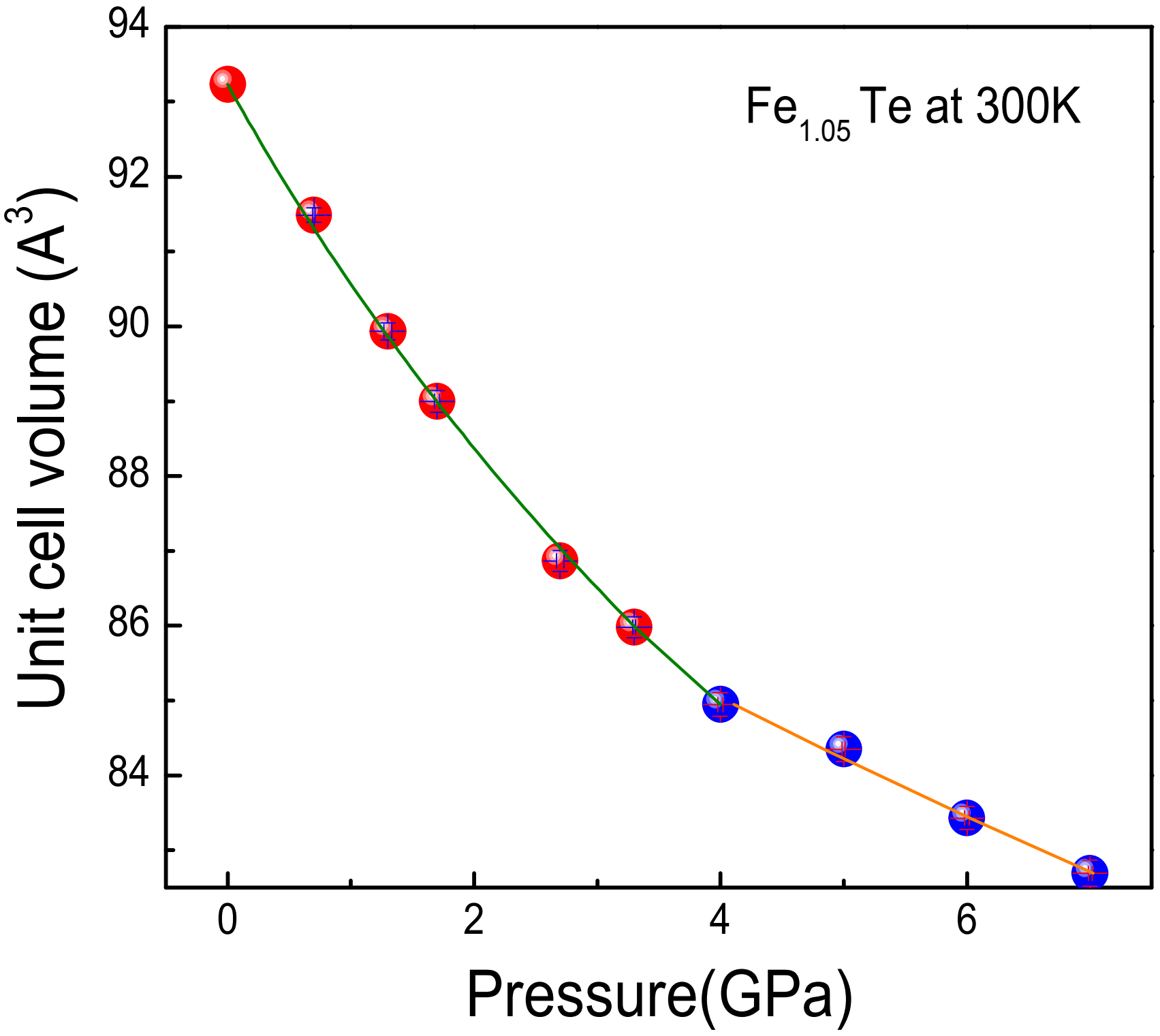

Fig.4

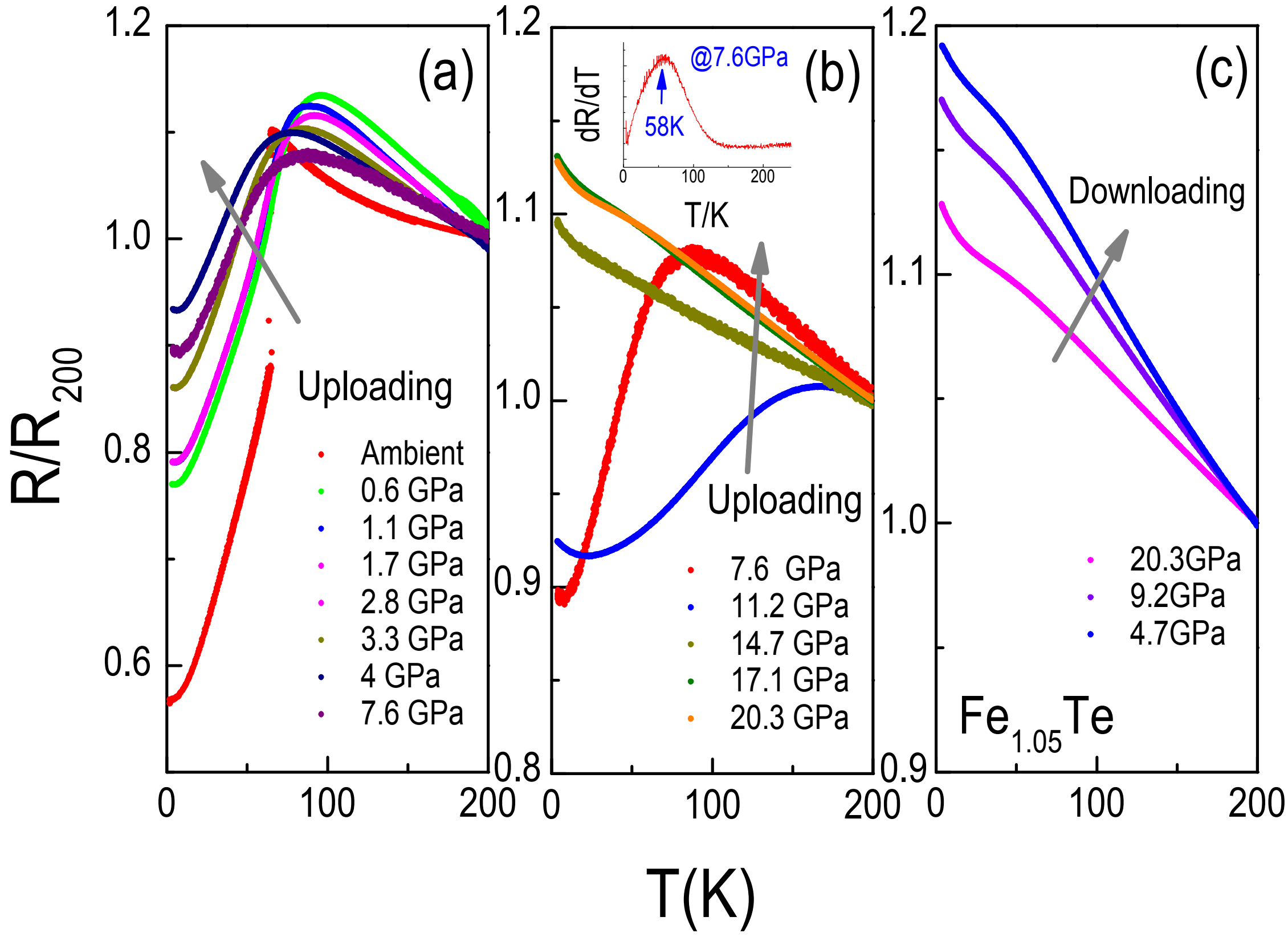

Fig. 5

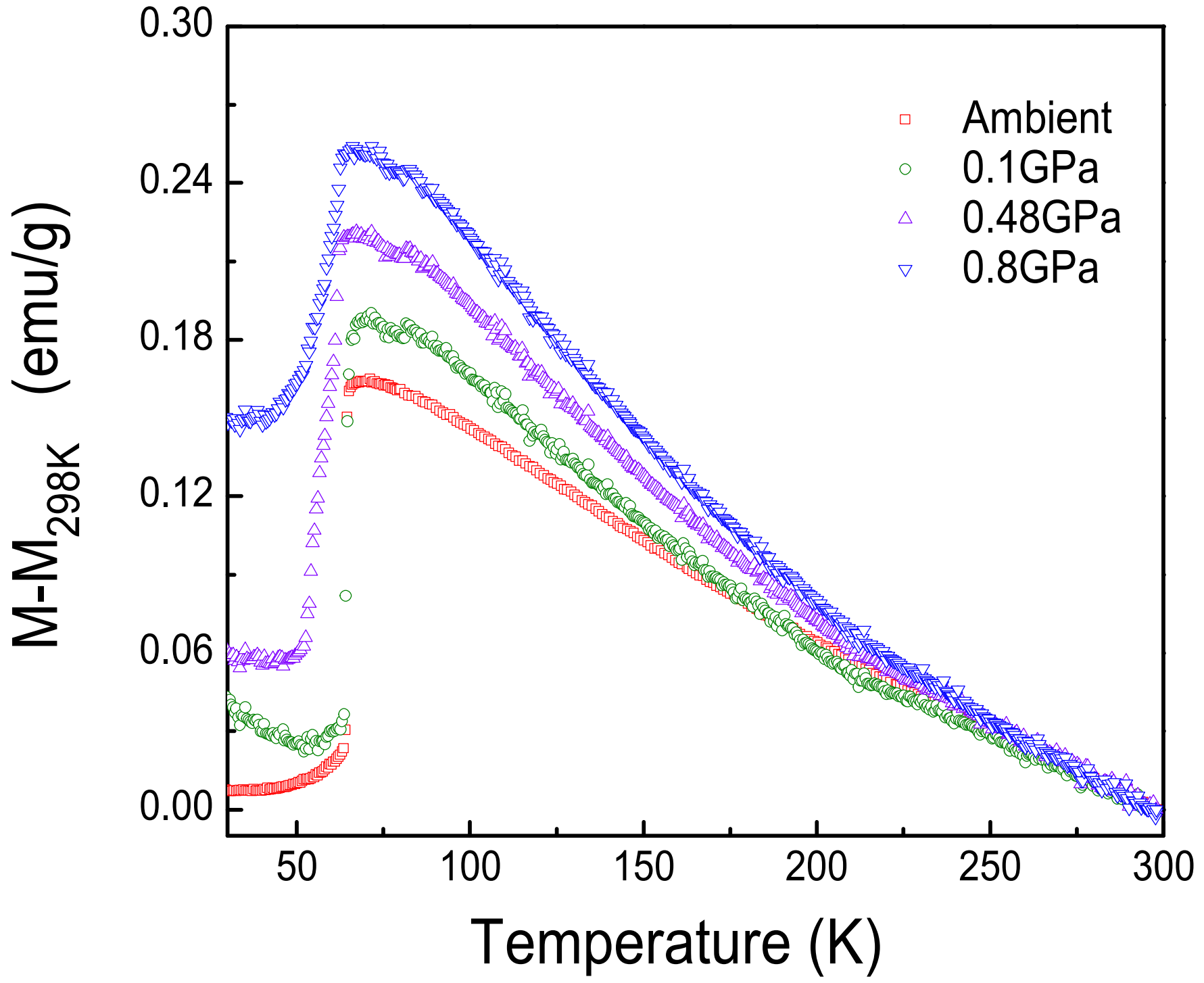

Fig. 6

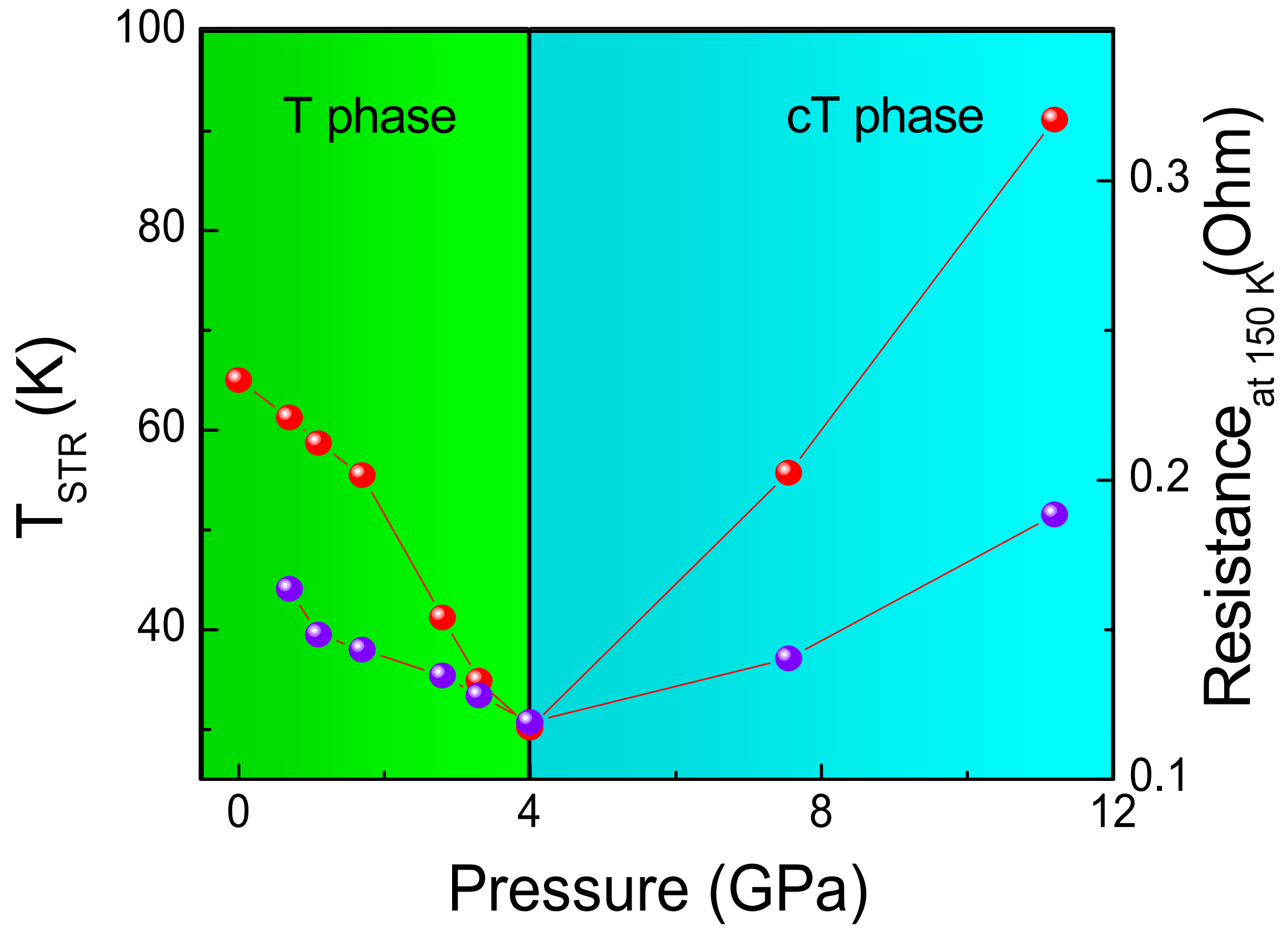

Fig.7

**Table 1** Bulk modulus $B_0$, volume at ambient pressure $V_0$ and pressure derivative $B_0'$ of tetragonal phase T and collapse tetragonal phase cT in $Fe_{1.05}Te$ compound.

| P (GPa) | Phase | $B_0$ | $V_0$(Å$^3$) | $B_0'$ |
|---|---|---|---|---|
| 0-4 | T | 31.3±1.45 | 93.231 | 6.6±1.2 |
| 4-7 | cT | 86.7±6.6 | 88.7±0.4 | 4 (fixed) |